\documentclass[final,3p,times]{elsarticle}

\usepackage{graphics}
\usepackage{graphicx}
\usepackage{epsfig}
\usepackage{amssymb}

\journal{Journal of Magnetism and Magnetic Materials}

\begin{document}
\begin{frontmatter}


\title{Thermal and magnetic phase transition properties of a binary alloy spherical nanoparticle:
A Monte Carlo simulation study}

\author[]{Z.D. Vatansever\corref{cor1}}
\cortext[cor1]{Corresponding author. Tel.: +90 3019531; fax: +90 2324534188.} \ead{zeynep.demir@deu.edu.tr}
\author[]{E. Vatansever}
\address{Department of Physics, Dokuz Eyl\"{u}l University, Tr-35160 \.{I}zmir, Turkey}



\begin{abstract}
We have used the Monte Carlo (MC) simulation method with Metropolis algorithm to study the finite temperature phase
transition properties of  a binary alloy spherical  nanoparticle with radius $r$ of the type $A_{p}B_{1-p}$.
The system consists of two different species of magnetic components, namely, $A$ and $B$, and  the components of the system
have been selected $A$ and $B$ to be as $\sigma = 1/2$ and $S=1$, respectively. A complete picture of phase
diagrams, total magnetizations and susceptibilities in related planes have been
presented, and the main roles of the radius of nanoparticle, active concentration value of type-$A$ atoms
as well as other system parameters on the thermal and magnetic phase transition features of the considered system
have been discussed in detail. Our MC investigations clearly show that  it is possible to
control the critical characteristic behaviors  of the system with the help of adjustable Hamiltonian parameters.

\end{abstract}

\begin{keyword}
Binary alloy systems, magnetic nanoparticles, Monte Carlo simulation.
\end{keyword}
\end{frontmatter}
\section{Introduction}\label{Introduction}
When the physical size of a magnetic system is reduced to a characteristic length, the system has
a bigger surface area to volume ratio  giving rise to a great many unusual thermal and magnetic
properties different from the conventional bulk systems \cite{Berkowitz}. In recent years,
there has been active research on small-size nanoparticles both experimentally and
theoretically because of their technological \cite{Plumer, Hayashi, Comstock} and  biomedical
applications \cite{Pankhurst, Rivas, Blakemore, Bazylinski, Ivanov}.
For example,  from the experimental point of view,  the multi-functionality nanowires
with an iron core and an iron oxide shell have been synthesized  by a facile low-cost fabrication
process  in Ref. \cite{Ivanov}. Here, Ivanov and co-workers report that a multi-domain state at remanence can
be obtained, which is an attractive feature for the biomedical applications.  Magnetic proximity effect
in ferrimagnetic/ferromagnetic core/shell Prussian blue analogues molecular magnet has been
found by Bhatt and co-workers  in Ref. \cite{Bhatt}. They synthesize  a ferrimagnetic core
of $\mathrm{Mn_{1.5}[Cr(CN)_{6}]\cdot7.5H_{2}O}$  surrounded by a ferromagnetic shell
of $\mathrm{Ni_{1.5}[Cr(CN)_{6}]\cdot7.5H_{2}O}$,  and note that such a process allows
us to enhance the critical temperature of the core/shell nanoparticle system, compared to the
bare-core and bare-shell critical temperatures.

On the other hand, from the theoretical point of view, a great deal of  studies have been devoted to
investigate the thermal and magnetic features of  nanoparticles such as nanocube, nanosphere,
nanorod, nanotube as well as nanowire by means of several types of methods such
as mean-field theory (MFT) \cite{Leite, Kaneyoshi1, Kaneyoshi2}, Green function
formalism (GF) \cite{Garanin}, cluster variation method (CVM) \cite{Wang}, effective-field theory (EFT)
with single-site correlations  \cite{Kocakaplan, Jiang1, Kaneyoshi3, Kaneyoshi4, Bouhou, Kantar, Hamri}, and
MC simulation  technique \cite{Leite, Magoussi, Margaris, Vasilakaki, Drissi, Zaim1, Zaim2, Zaim3,
Aouini, Feraoun, Jiang2, Russier}. For example, by making use of MC simulation technique with Metropolis
algorithm, thermal and magnetic phase transition properties of antiferromagnetic/ferrimagnetic core/shell
nanoparticles have been studied by Vasilakaki and co-workers in detail \cite{Vasilakaki}.
It has been reported that the coercivity and loop shift show a non-monotonic dependence with the core diameter
and shell thickness, in agreement with the experimental data. Magnetic phase transitions as well
as hysteretic features of a graphyne core/shell nanoparticles have been investigated in Ref. \cite{Drissi} based on
MC simulation method. It has been found that the considered system exhibits a number of unusual and
characteristic treatments such as the occurrence of one and two compensation temperatures. Moreover,
magnetic phase transition properties of a single spherical core/shell
nanoparticle, consisting of a ferromagnetic core surrounded by a ferromagnetic shell with antiferromagnetic
interface coupling have been realized within the framework of MC simulation scheme \cite{Zaim1}.
It is underlined that for appropriate values of the system parameters, multiple compensation points
may emerge in the system. They also analyzed the core radius and shell thickness dependencies of the
critical behavior of the system, and they found that the values of compensation and critical temperatures vary with
changing value of the particle size, and they reach saturation values for high values of the
particle size.

Furthermore, determination of equilibrium phase transition properties of binary alloy systems containing
disorder effects problems which may be arisen from a random distributions of the magnetic
components or random exchange interactions between the  magnetic components in the material
has a long history. Many studies have been performed regarding the physical properties of
disordered binary magnetic materials with quenched randomness where the random  variables of
a system may not change its value over time based on a variety of techniques
such as MFT \cite{Thorpe, Tahir, Plascak, Katsura}, EFT \cite{Honmura, Kaneyoshi5, Kaneyoshi6,
Kaneyoshi7, Kaneyoshi8} as well as MC \cite{Scholten1, Scholten2, Godoy, Cambui}, in the context of
bulk materials.  We learned  from these works  quenched disordered binary alloy systems with different signs and
unequal magnitudes of the spin-spin interactions and single ion-anisotropy exhibit unusual and
interesting thermal and magnetic behaviors such as presence  of a reentrant type character in
the magnetization versus temperature profile, compensation and also spin-glass behaviors. Keeping the
discussions mentioned above in mind,  we intend to elucidate  the finite temperature phase transitions
and critical properties of a binary alloy spherical nanoparticle with radius $r$ of the type $A_{p}B_{1-p}$ in this study.
The system consists of two different species of magnetic spin components, i.e., $A$ and $B$, and we
select the components of the system $A$ and $B$ to be as $\sigma = 1/2$ and $S=1$, respectively. In the system,
there aren't any un-occupied lattice sites and each lattice site is occupied by type-$A$ or $-B$ atom,
depending on the active concentration value of type-$A$ components, $p$.
We perform MC simulation, using Metropolis algorithm and determine the effects of the
$p$, radius of spherical nanoparticle as well as other system parameters
on the phase transition features of the considered system. To the best our knowledge, there are no
works regarding the thermal phase transition properties of binary alloy type spherical nanoparticle systems, except from
Ref. \cite{Zaim3} where compensation behavior of a ferrimagnetic spherical nanoparticle system with binary alloy shell
is studied by means of MC simulation method. It has been stated that  the system exhibits one,
two or even three compensation points, depending on the system parameters. We should note that our model is completely
different from the system studied in Ref. \cite{Zaim3}. That is to say,
the magnetic components $A$ and $B$ can locate  any place of the nanoparticle, whereas they can only
locate the shell part of the particle, in the mentioned work.

The plan of the remainder parts of the paper is as follows: In section \ref{Formulation},
we present our model. The results and discussions are given in section \ref{Results}, and finally
section \ref{Conclusion} contains our conclusions.

\section{Formulation}\label{Formulation}
\begin{figure}[!h]
\begin{center}
\includegraphics[width=3.15cm,height=2.5cm]{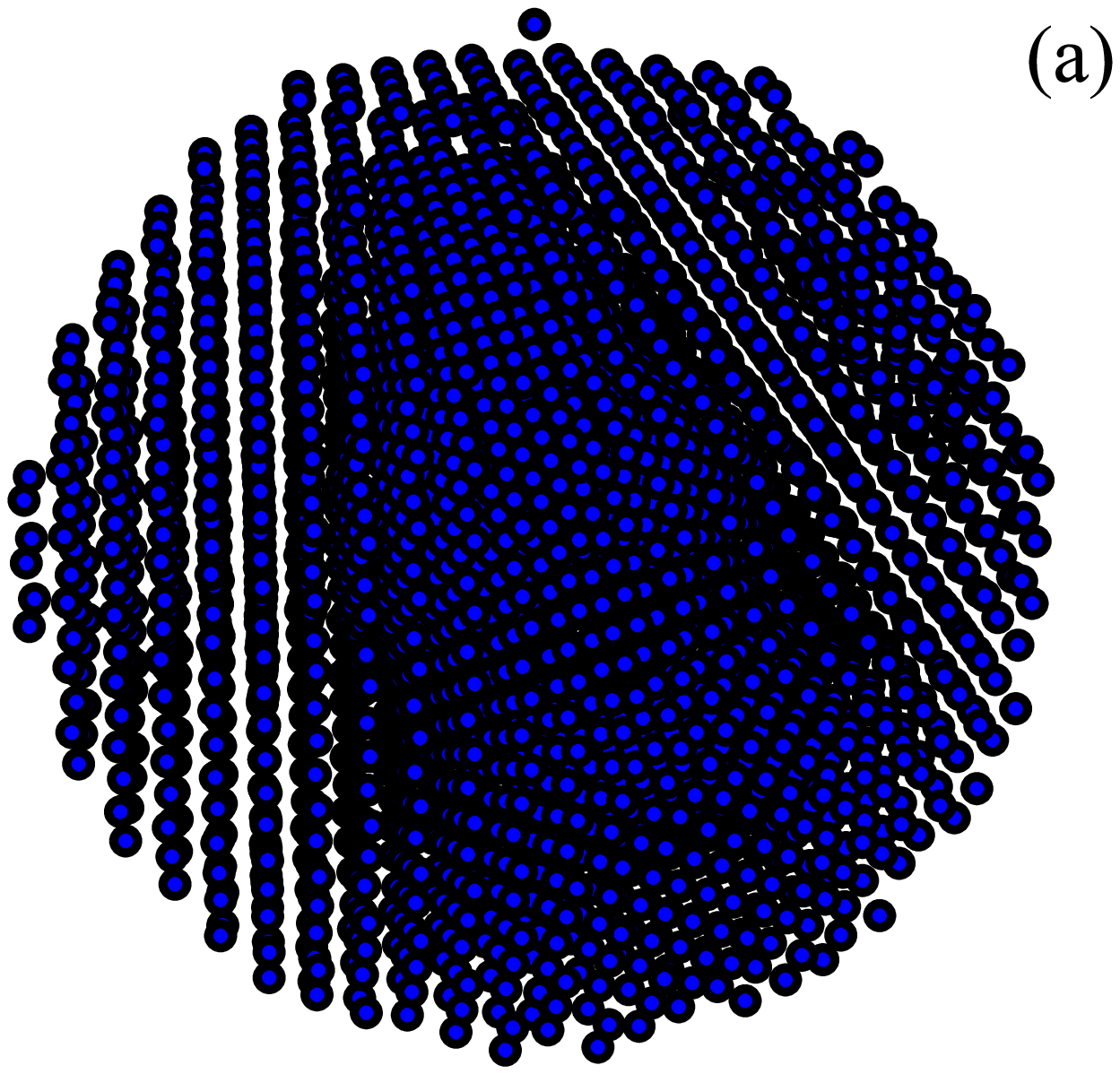}
\includegraphics[width=3.15cm,height=2.5cm]{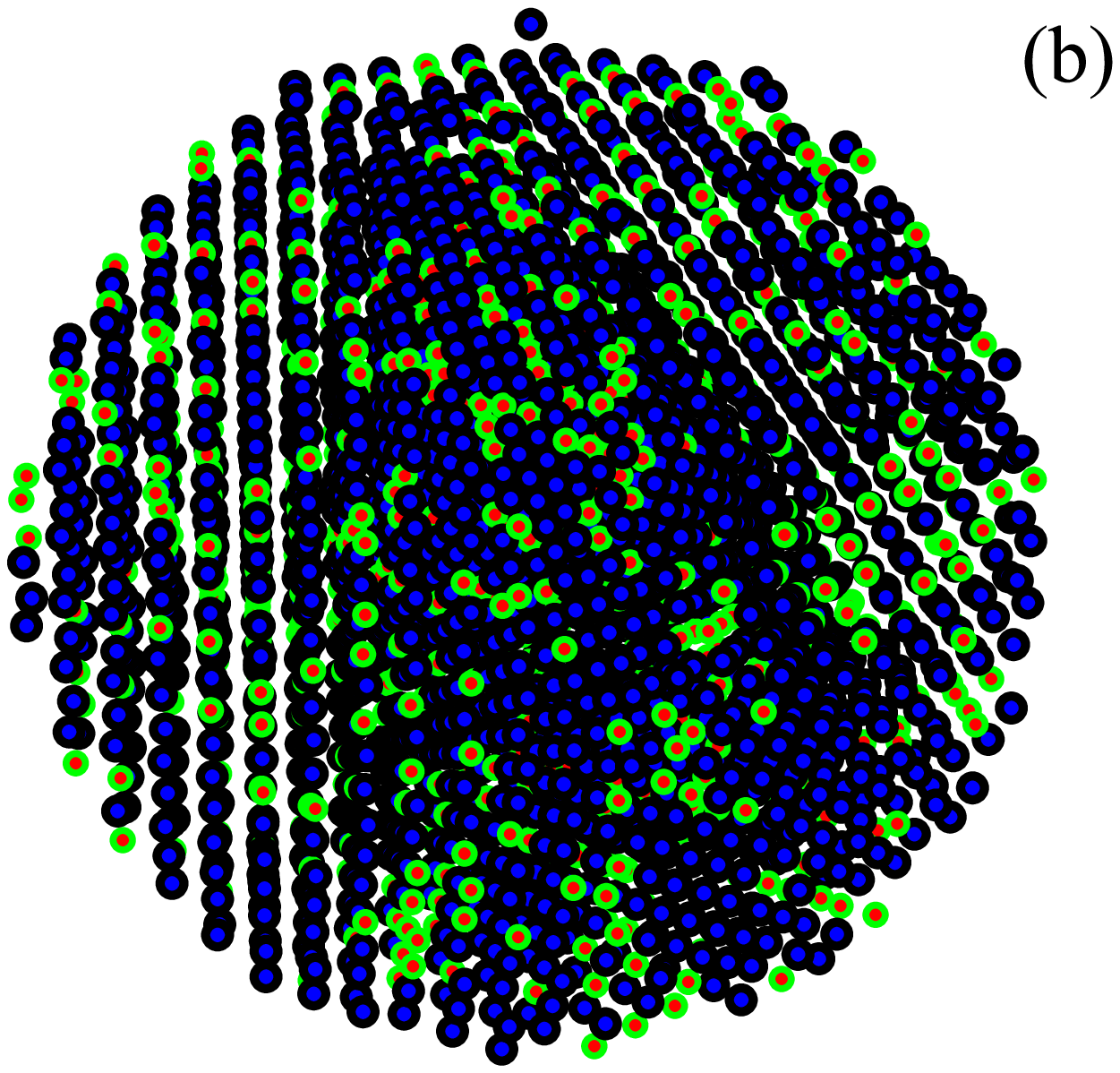}
\includegraphics[width=3.15cm,height=2.5cm]{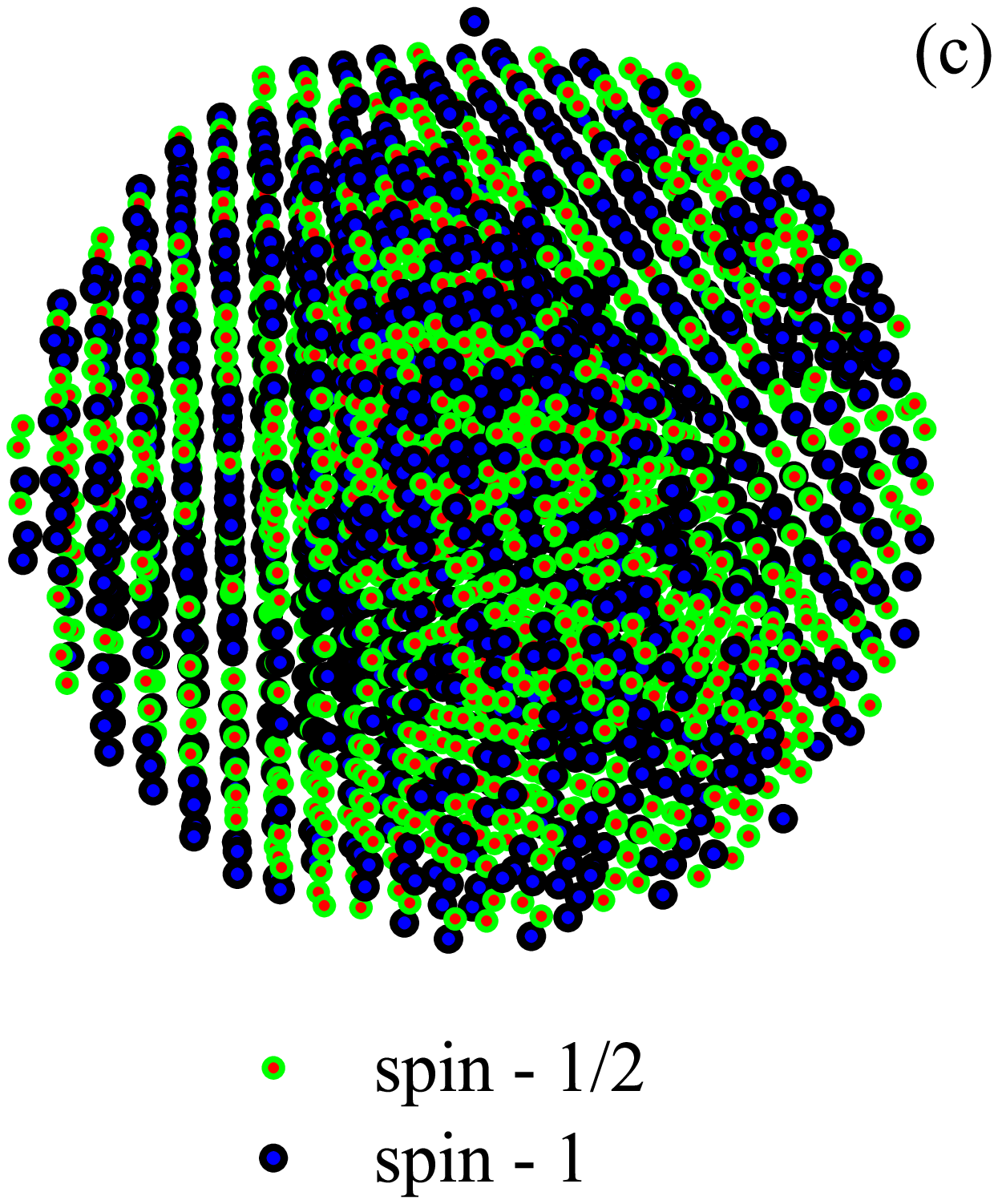}
\includegraphics[width=3.15cm,height=2.5cm]{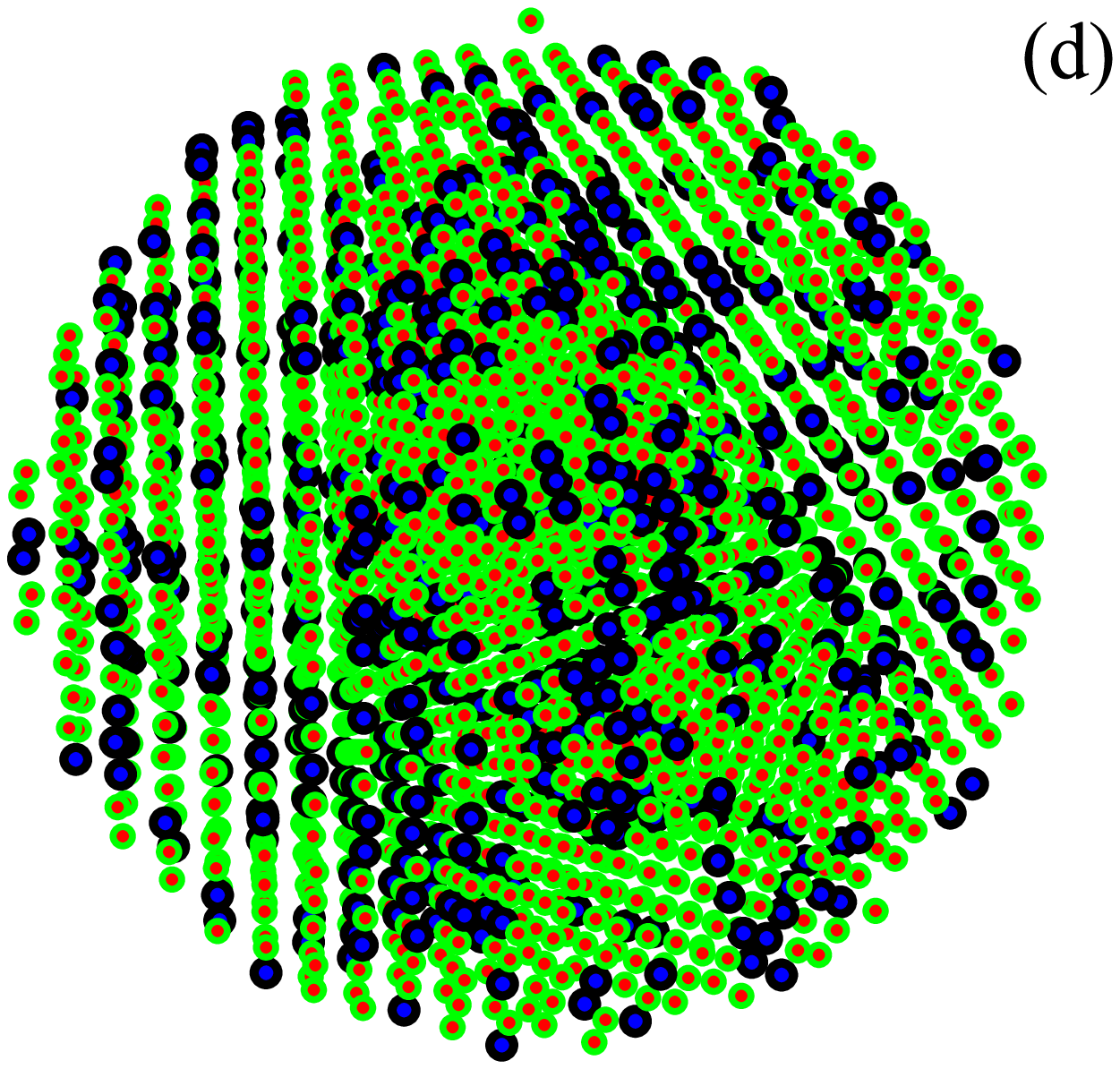}
\includegraphics[width=3.15cm,height=2.5cm]{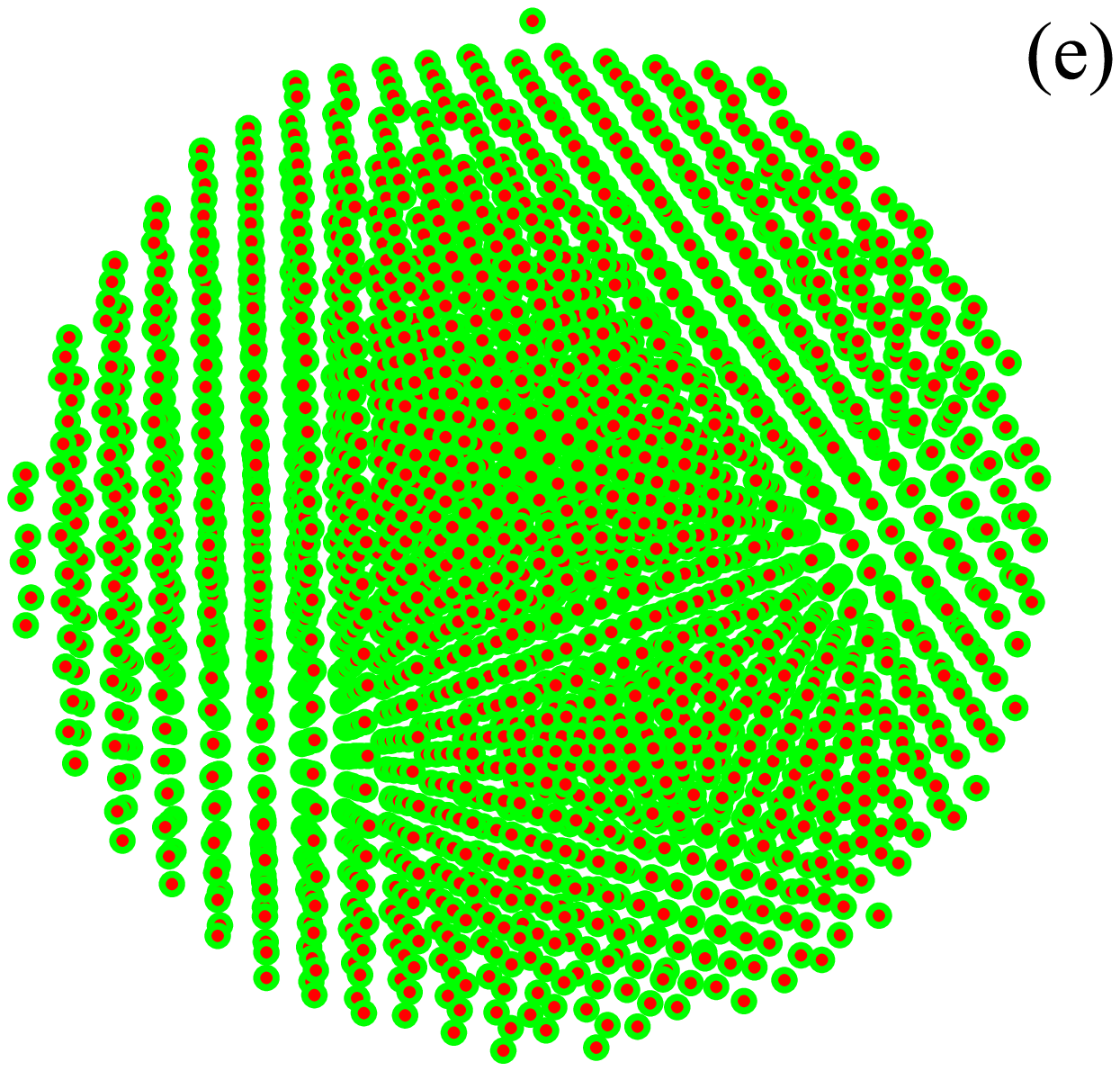}
\caption{(Color online) Schematic representations of the binary alloy spherical nanoparticle of the type $A_{p}B_{1-p}$
for five selected active concentration values of magnetic components, namely (a) $p=0$,  (b) $p=0.25$,
(c) $p=0.5$, (d) $p=0.75$ and (e) $p=1$. }\label{Fig1}
\end{center}
\end{figure}

We consider a binary alloy spherical nanoparticle system of the type $A_{p}B_{1-p}$ with total radius $r$,
which is schematically shown in figure \ref{Fig1}. The lattice sites are randomly occupied by two
different species of magnetic components $A$ and $B$ with the
concentration $p$ and $1-p$, respectively.   The Hamiltonian describing our model of magnetic system is given by:
\begin{equation}\label{Eq1}
\hat{H}=-J\sum_{\langle i,j \rangle}\left[\delta_{iA}\delta_{jA}\sigma_{i}\sigma_{j}+
\delta_{iB}\delta_{jB}S_{i}S_{j}+
\delta_{iA}\delta_{jB}\sigma_{i}S_{j}+
\delta_{iB}\delta_{jA}S_{i}\sigma_{j}\right]-\Delta\sum_{i}\delta_{iB}S_{i}^2,
\end{equation}
here $J>0$ denotes the ferromagnetic spin-spin interaction term between nearest neighbor spins while $\Delta$ refers to
the single-ion anisotropy term. The $\sigma$ and $S$ are  classic Ising spin variables which can take
values of $\sigma = \pm 1/2$ and $S=\pm 1, 0$ for the magnetic components-$A$ and $-B$ of system, respectively.
The symbol $\delta_{i\alpha}=1$ ($\alpha= A$ or $B$) if site $i$ is occupied by type-$\alpha$ atom and $0$ otherwise.
The first summation in Eq. (\ref{Eq1}) is over the nearest neighbor pairs while the second one is over all lattice
sites occupied by type-$B$ atoms.

In order to investigate the thermal and magnetic phase transition properties of the binary alloy type spherical
nanoparticle system, we use the MC simulation technique with local spin update Metropolis
algorithm \cite{Binder, Newman}. We simulate the system with radius $r$ up to $18$, located on a simple
cubic lattice, under free boundary conditions applied in all directions. We can summarize the simulation
procedure as follows. First, the simulation  starts at a high temperature value using random initial condition,
and then the system is slowly cooled down with a reduced temperature step $k_{B}\Delta T/J=2\times10^{-2}$, where
$k_{B}$ and $T$ are Boltzmann constant and absolute temperature, respectively. The spin configurations are generated
by selecting the sites sequentially through the binary alloy type nanoparticle system, and making single-spin flip
attempts, which are accepted/rejected according to the Metropolis algorithm. Thermal variations of various thermodynamic
quantities are generated over $50$ independent computer experiments. In each computer experiment, the first
$10^4$ MC steps have been discarded for thermalization process, and the numerical data are collected over the
next $4\times10^4$ MC steps. Based on our test investigations, we  note that  this amount of transient steps are
found to be sufficient for thermalization for the whole range of the parameter sets.

Our program calculates the the instantaneous values of the magnetizations as follows:
\begin{equation}\label{Eq2}
M_{A}=\frac{1}{N_{A}}\sum_{i=1}^{N_{A}}\sigma_{i}, \quad \quad \quad M_{B}=\frac{1}{N_{B}}\sum_{i=1}^{N_{B}}S_{i}, \quad \quad \quad
M_{T}=\frac{N_{A}M_{A}+N_{B}M_{B}}{N_{A}+N_{B}}
\end{equation}
where $N_{A}$ and $N_{B}$ correspond to the total number of ions $A$ and $B$ in the system, respectively. In order to determine
the magnetic phase transition point separating the ordered and disordered phases from each other,
we use and check the thermal variation of the total susceptibility, which is defined by:
\begin{equation}\label{Eq3}
\chi_{T}=(N_{A}+N_{B})\frac{\left(\langle M_{T}^2\rangle -\langle M_{T} \rangle^2 \right)}{k_{B}T}.
\end{equation}

\section{Results and Discussion}\label{Results}
In this section, we present our MC simulation results of a binary alloy nanoparticle with a spherical shape.
We discuss the magnetic properties and critical behaviour of the system by analysing the phase diagrams in various planes.

\begin{figure*}[!h]
\center
\includegraphics[width=4.97cm]{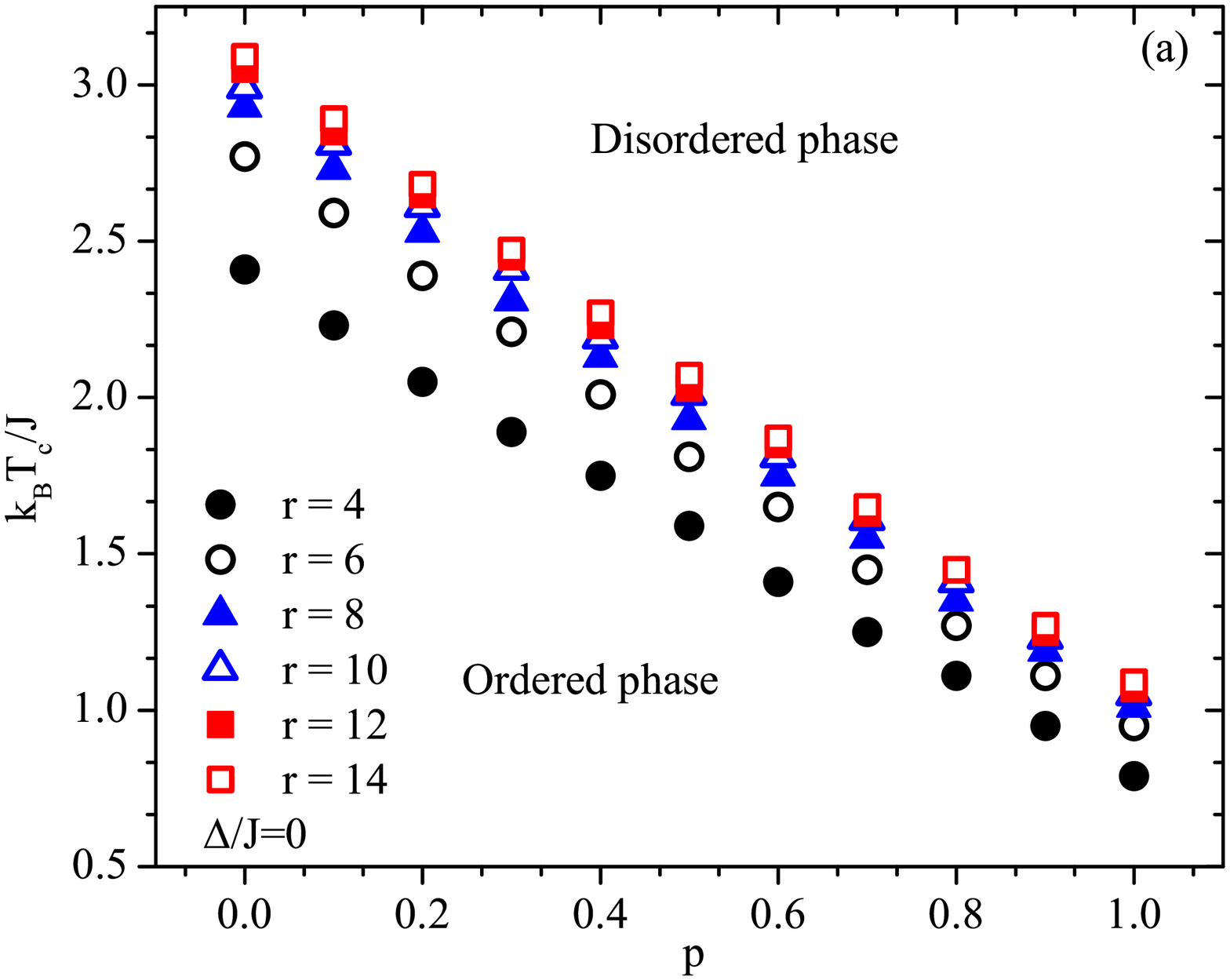}
\includegraphics[width=5.cm]{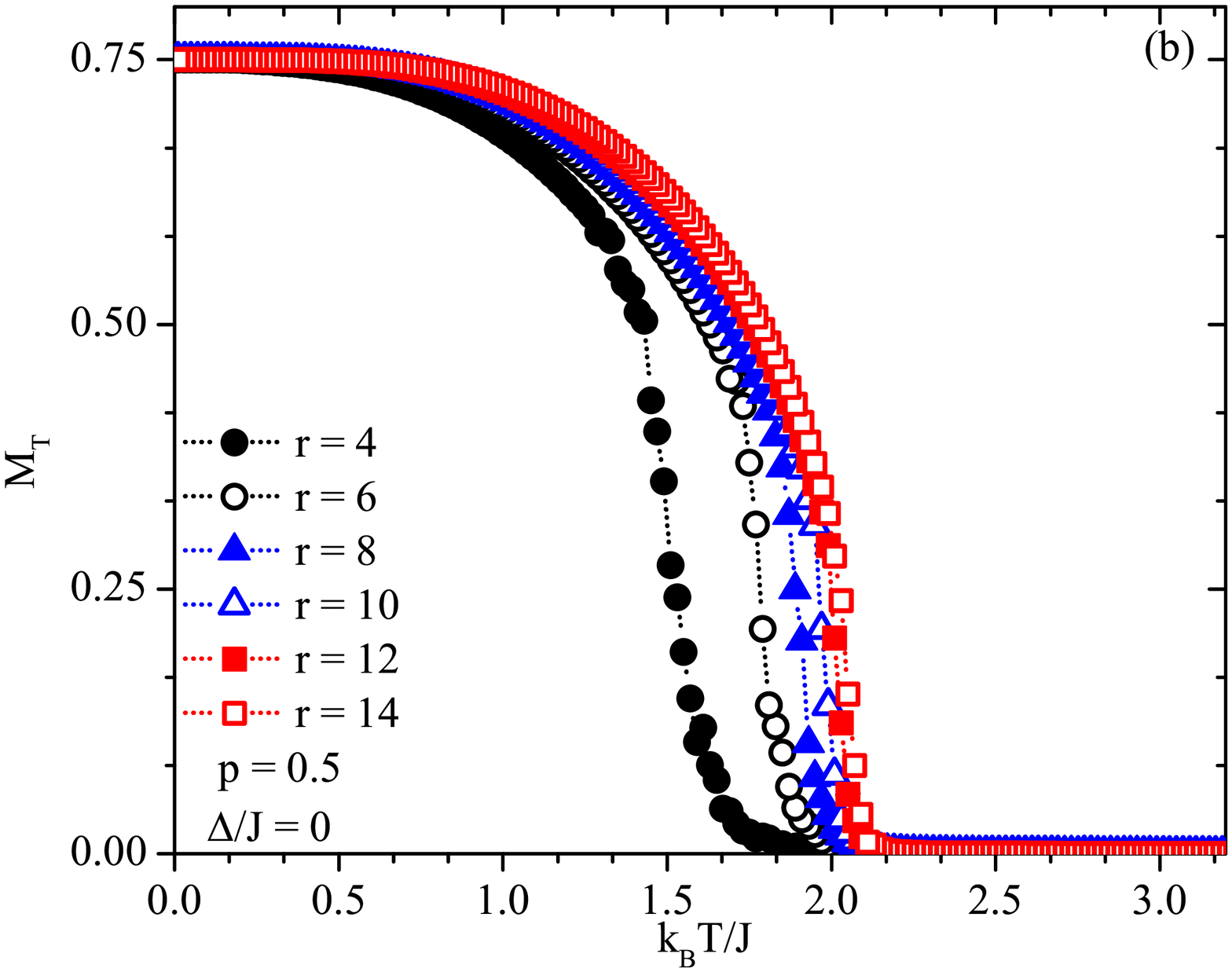}
\includegraphics[width=5.cm]{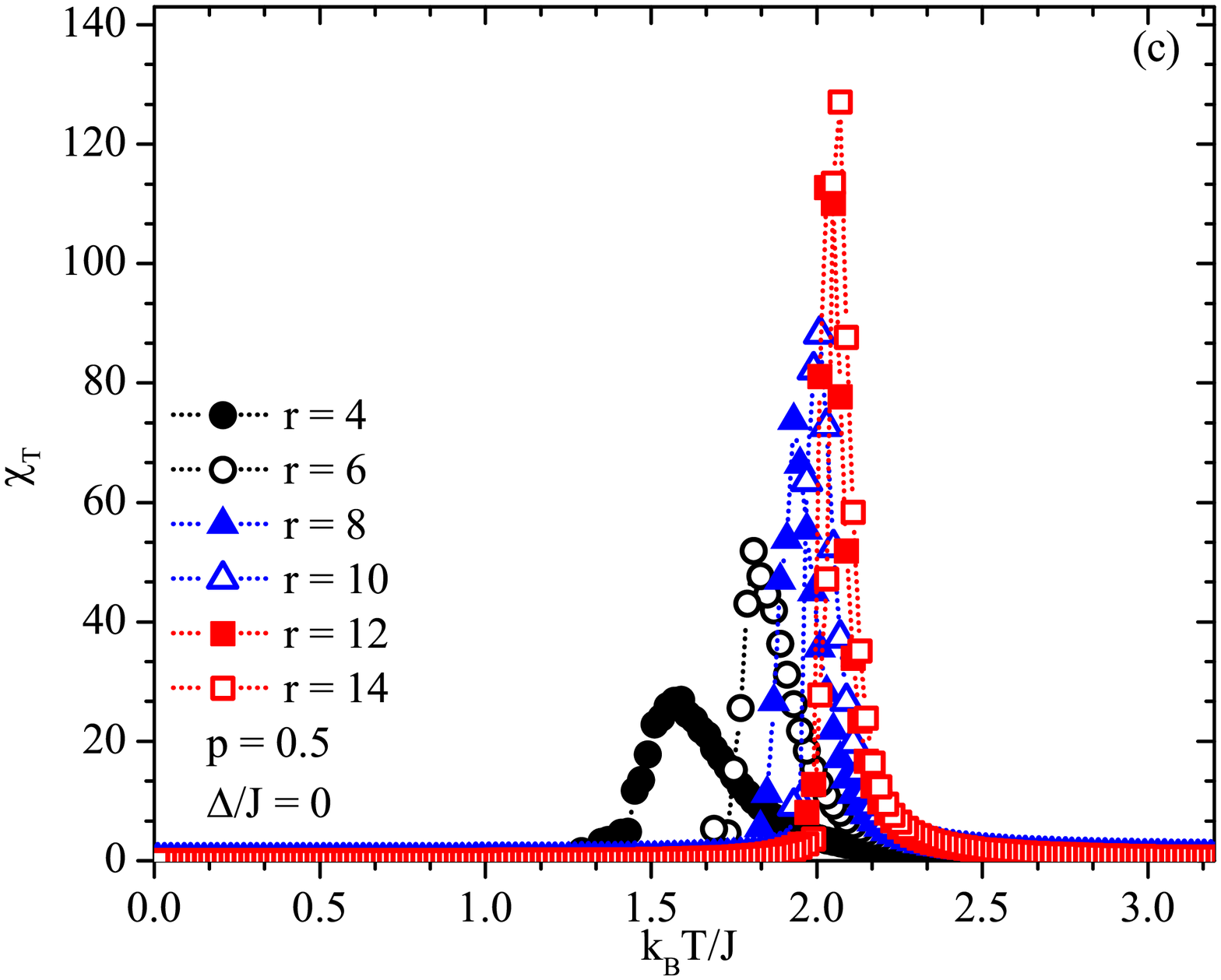}

\caption{(Color online) (a) Magnetic phase diagrams in $(p-k_{B}T_{C}/J)$ plane  of the binary alloy
spherical nanoparticle system. The curves are presented for various values of radius of the
spherical nanoparticle,  $r= 4, 6, 8, 10, 12$ and $14$.  Effects of the radius on the
thermal dependencies of total (b) magnetization $M_{T}$ and (c) susceptibility $\chi_{T}$ of the system
for $p=0.5$. All figures are obtained for value of  $\Delta/J=0$. }\label{Fig2}
\end{figure*}

Firstly, we investigate the dependence of the transition temperature from ferromagnetic to paramagnetic phase
of the binary alloy nanoparticle on the concentration of spin-$1/2$ atoms in the absence of crystal-field
interaction. Figure \ref{Fig2}(a) shows the phase-diagram of the system in $(p-k_{B}T_{C}/J)$ plane for
six selected values of nanoparticle radius $r= 4, 6, 8, 10, 12$ and $14$. From the figure one can observe
that for all the values of nanoparticle radius under consideration, as the number of type-$A$ atoms is increased
starting from  $p=0$ to $1$, the transition point moves to a lower value in the temperature axis
and thereby ordered regions in $(p-k_{B}T_{C}/J)$ becomes narrower.
This is due to the fact that when one increases $p$, the energy contribution coming from the exchange interaction
term becomes smaller and consequently the system exhibits phase transition at lower temperature values.

\noindent A second point which should be noted about Figure \ref{Fig2}(a) is the
variation of the phase diagram separating  the spin-ordered and disordered phases with
the size of the nanoparticle. It is obvious that when the radius  of the nanoparticle
begins to rise the critical temperature of the system starts to shift to  higher
temperature values for all $p$. This can be observed explicitly
from Figure \ref{Fig2}(b) and \ref{Fig2}(c) where the total magnetization $M_{T}$ and susceptibility $\chi_{T}$ of
the system are depicted as a function of the temperature with varying particle radius for $p=0.5$. It is seen that
the total magnetization of the system gradually decreases starting from its saturation value of $M_{T}=0.75$ with increasing
thermal agitation, and  it vanishes continuously at different critical temperature values
depending on the considered value of  nanoparticle size. As seen from the Figure \ref{Fig2}(c), the
susceptibility curves demonstrate a smooth cusp for smaller binary alloy spherical nanoparticles. When the
radius of the nanoparticle increases, a divergent treatment emerges in the susceptibility curve
at the phase transition point which indicates  of a second-order phase transition. In addition, the
position of the susceptibility peak moves to higher temperature region  with increasing
nanoparticle radius. It should be emphasized here that our numerical findings regarding the thermal and
phase transition properties of the present system are in accordance with the
recently published works where equilibrium and non-equilibrium phase transition
features of different types of binary alloy  systems \cite{Godoy, Cambui} as well as a clean nanowire with radius $r$ and length $L$
under the influence of a time dependent magnetic field have been implemented by utilizing MC simulation technique \cite{Yuksel}.

\begin{figure*}[!h]
\center
\includegraphics[width=4.97cm]{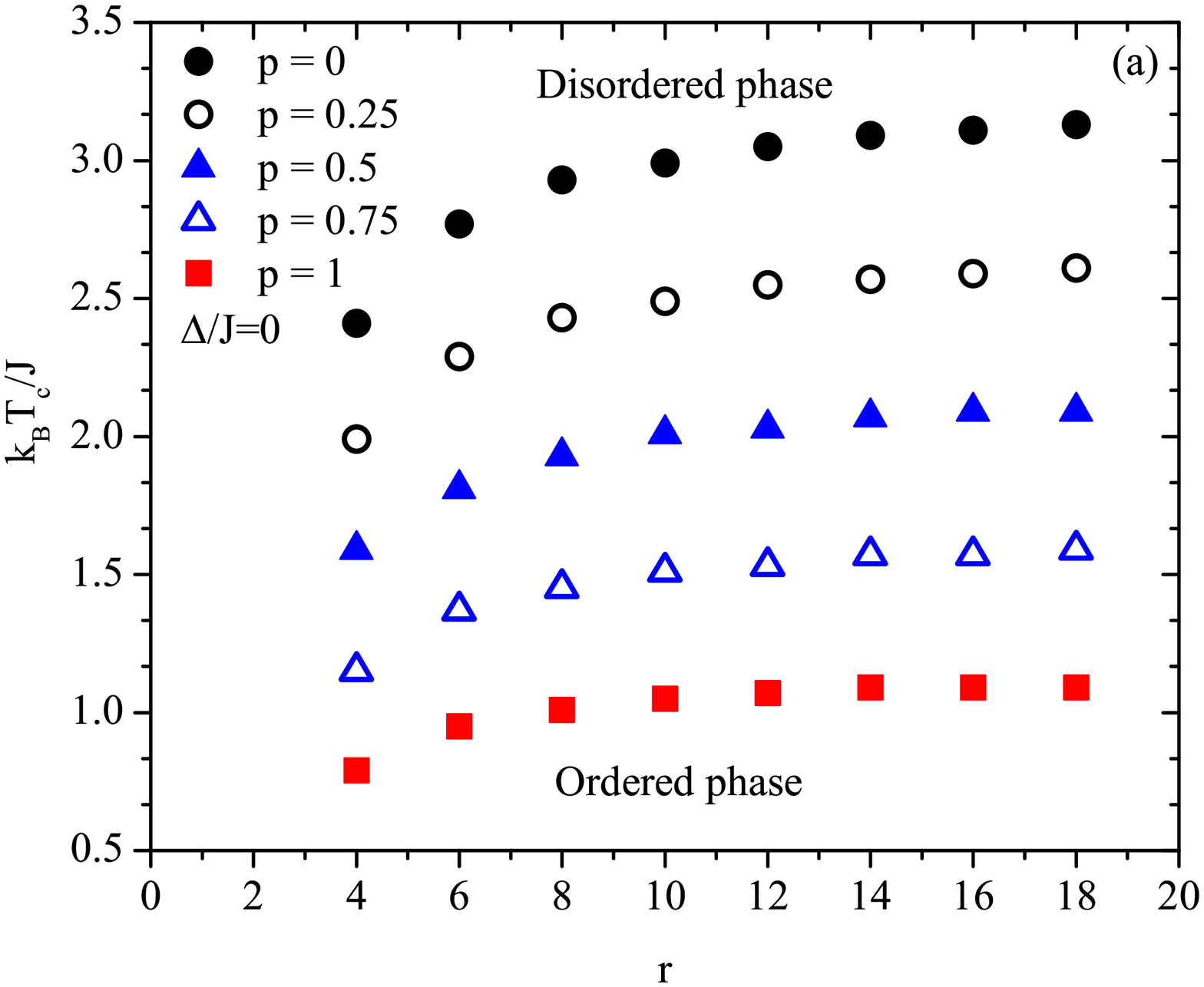}
\includegraphics[width=5.cm]{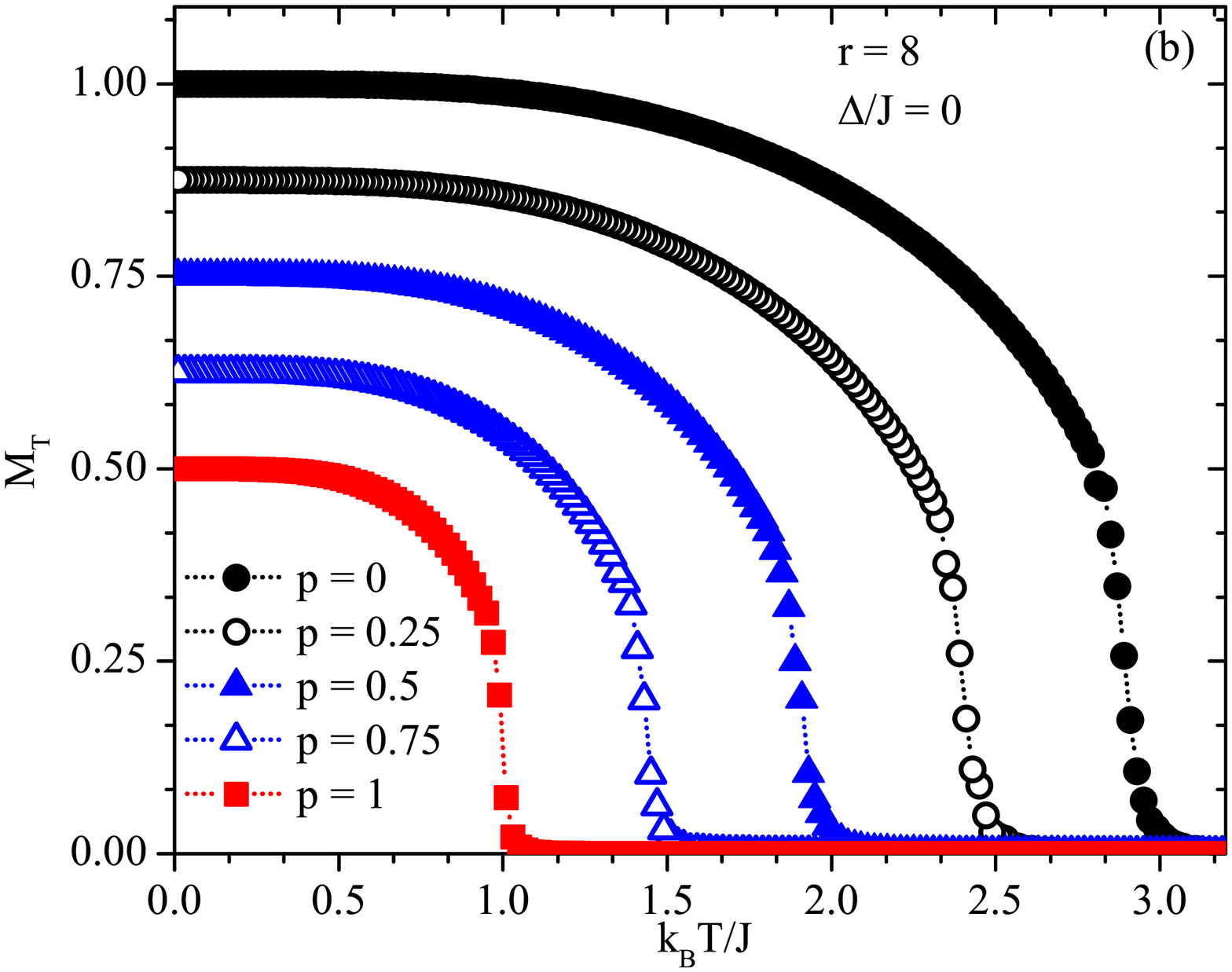}
\includegraphics[width=5.cm]{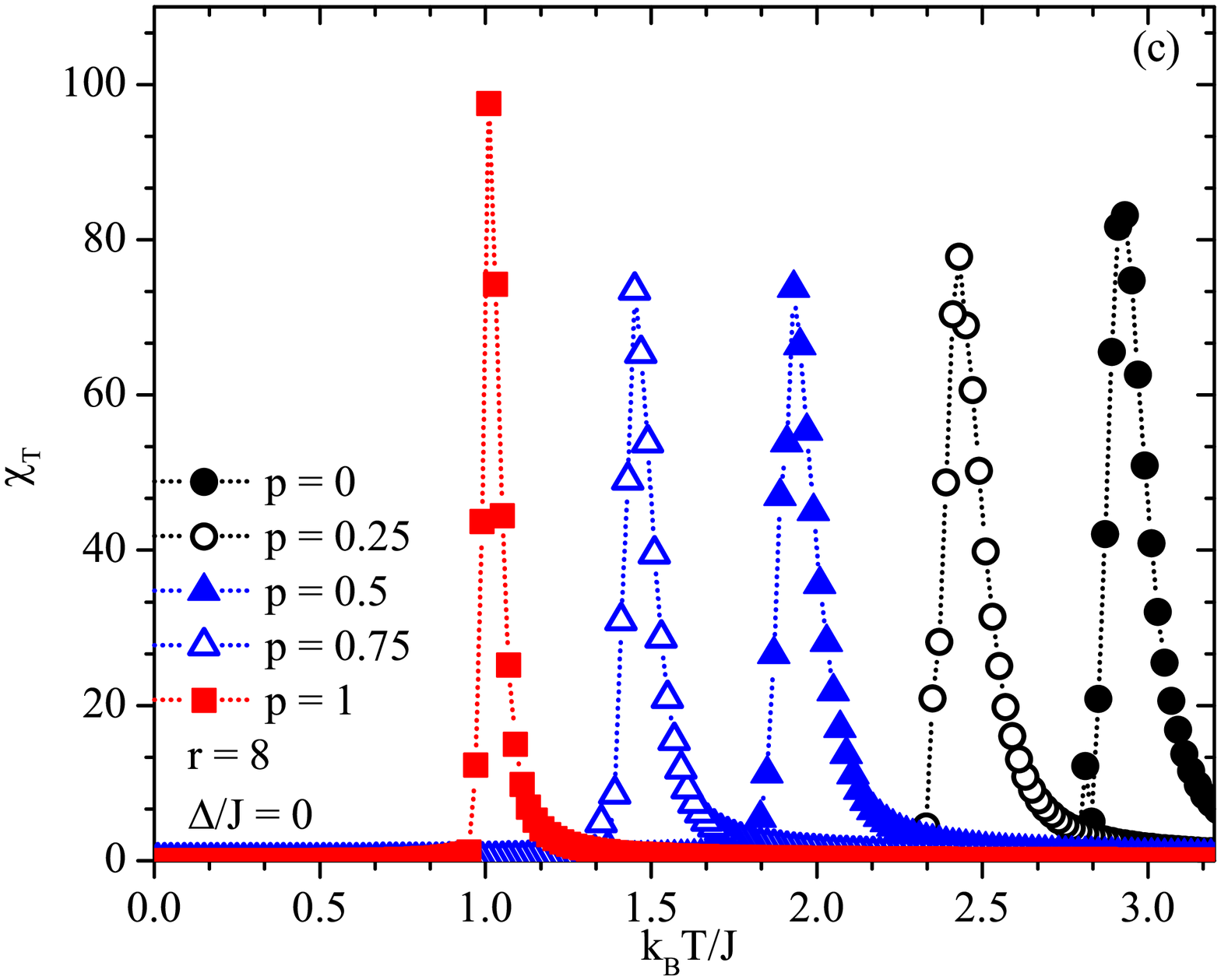}
\caption{(Color online) Radius dependency  of the critical point of binary alloy spherical
nanoparticle system for considered values of active concentrations of type-$A$ atoms
such as $p=0, 0.25, 0.5, 0.75$ and $1$. Influences of the active concentration of
type-$A$ atoms on the temperature dependencies of total (b) magnetizations $M_{T}$ and
(c) susceptibility $X_{T}$ of the system with $r=8$ corresponding to the
Figure \ref{Fig3}(a). All figures are displayed  for
value of $\Delta/J=0$.}\label{Fig3}
\end{figure*}

Next, we examine the change of the transition temperature from ordered to disordered phase with the nanoparticle
radius $r$. Figure \ref{Fig3}(a) shows the critical temperature as a function of the nanoparticle radius for
different values of the concentration of type-$A$ atoms in the system in the case of $\Delta/J=0$.
Due to the high surface to volume ratio for small size particle, and the lower coordination number of the
surface atoms, the phase transition temperature of the system decreases with the reduction of its size.
This fact is evidently seen in Figure \ref{Fig3}(a). As the size of the system is increased
starting from $r=4$, the phase transition temperature moves to higher temperature value. The reason  is that when the size
of the nanoparticle becomes larger, the energy contribution from the spin-spin interaction
term increases and thus more thermal energy is required for occurrence of  a phase transition.
As one further  rises $r$, the critical temperature becomes nearly independent of the
particle-size, where surface effects may be negligible,  and it saturates at a  certain temperature
value which sensitively depends on the concentration of type-$A$ atoms. We show the thermal dependencies of the total
magnetization and susceptibility curves for several values of concentration of type-$A$ atoms and a fixed value of
radius $r=8$ of particle in Figure \ref{Fig3}(b) and \ref{Fig3}(c) corresponding to the Figure \ref{Fig3}(a),
respectively.  The behaviour of the total magnetization of the binary alloy nanoparticle system
exhibits a second-order phase transition at different critical temperature
values depending on $p$. Also the positions of the peaks of the susceptibility curves support the magnetization data
and shift to lower temperature region with increasing $p$. It should be pointed out that
melting temperature dependencies of the various types of binary alloy nanoparticle, for
example $\mathrm{In_{0.3}Sn_{0.7}}$, $\mathrm{CuNi}$ and $\mathrm{Cu_{0.25}Ni_{0.75}}$ have been discussed
in Ref. \cite{YunBin}, in the  context of size and composition of binary nanoparticles.

\begin{figure*}[!h]
\center
\includegraphics[width=4.9cm]{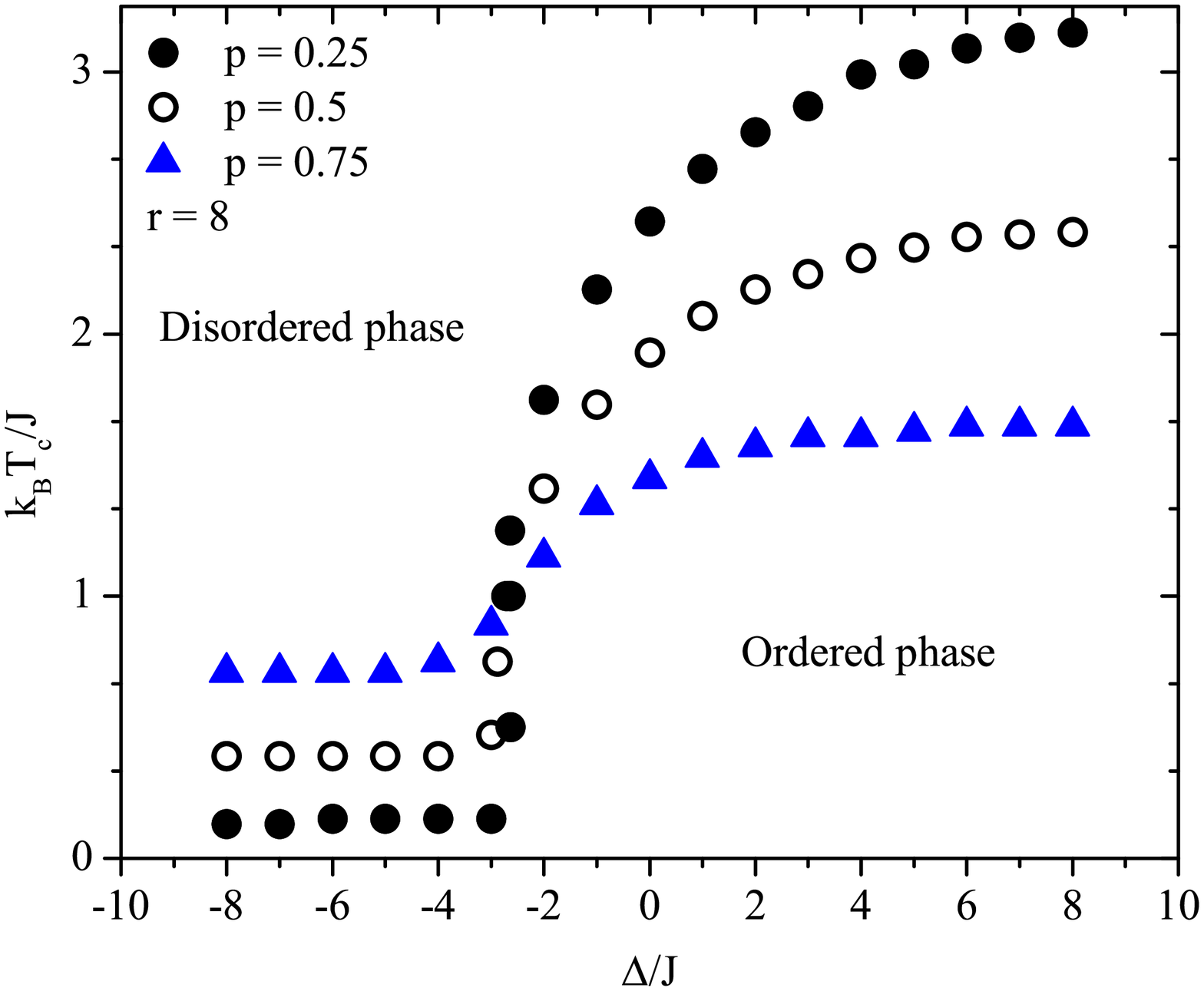}
\includegraphics[width=5.cm]{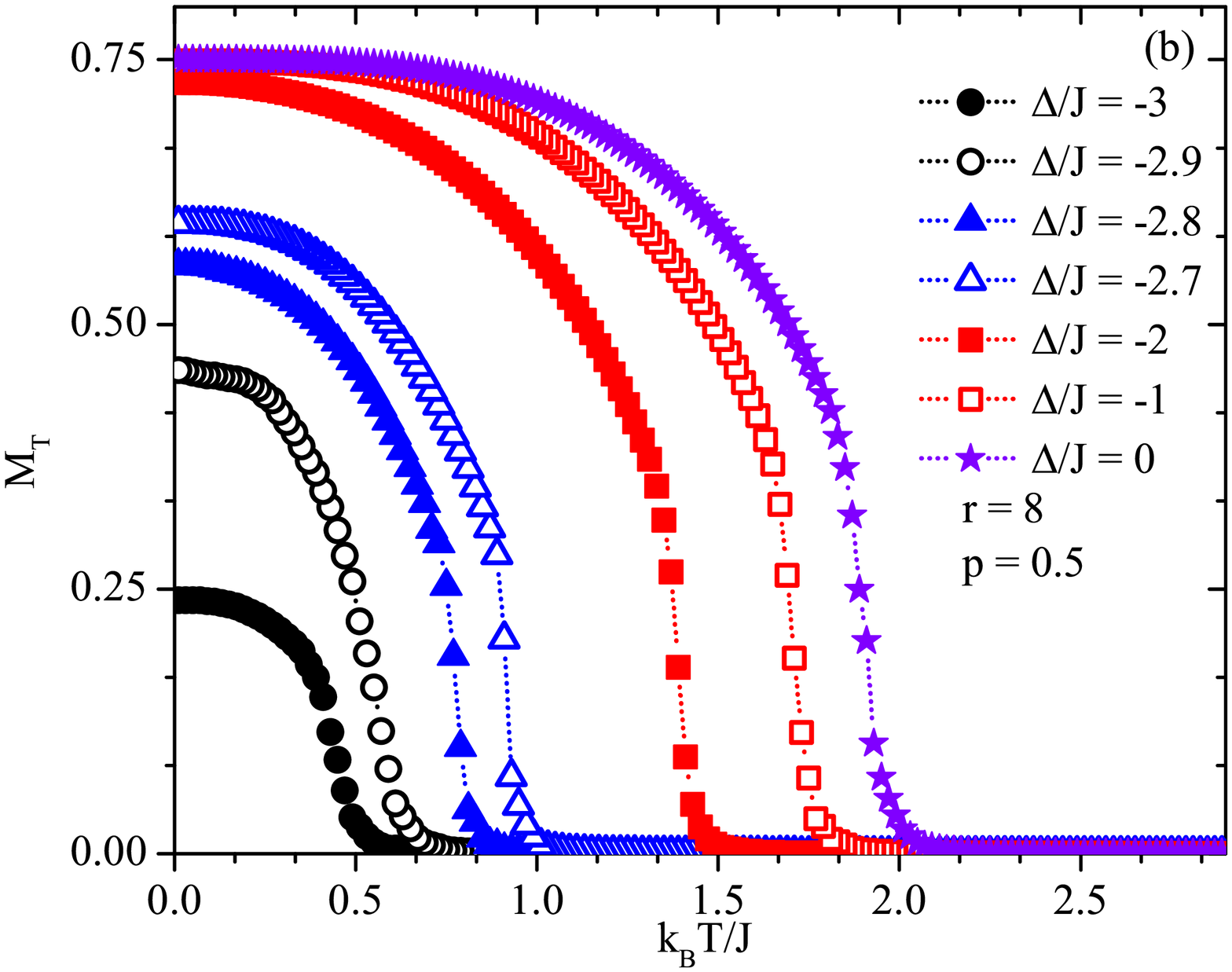}
\includegraphics[width=5.cm]{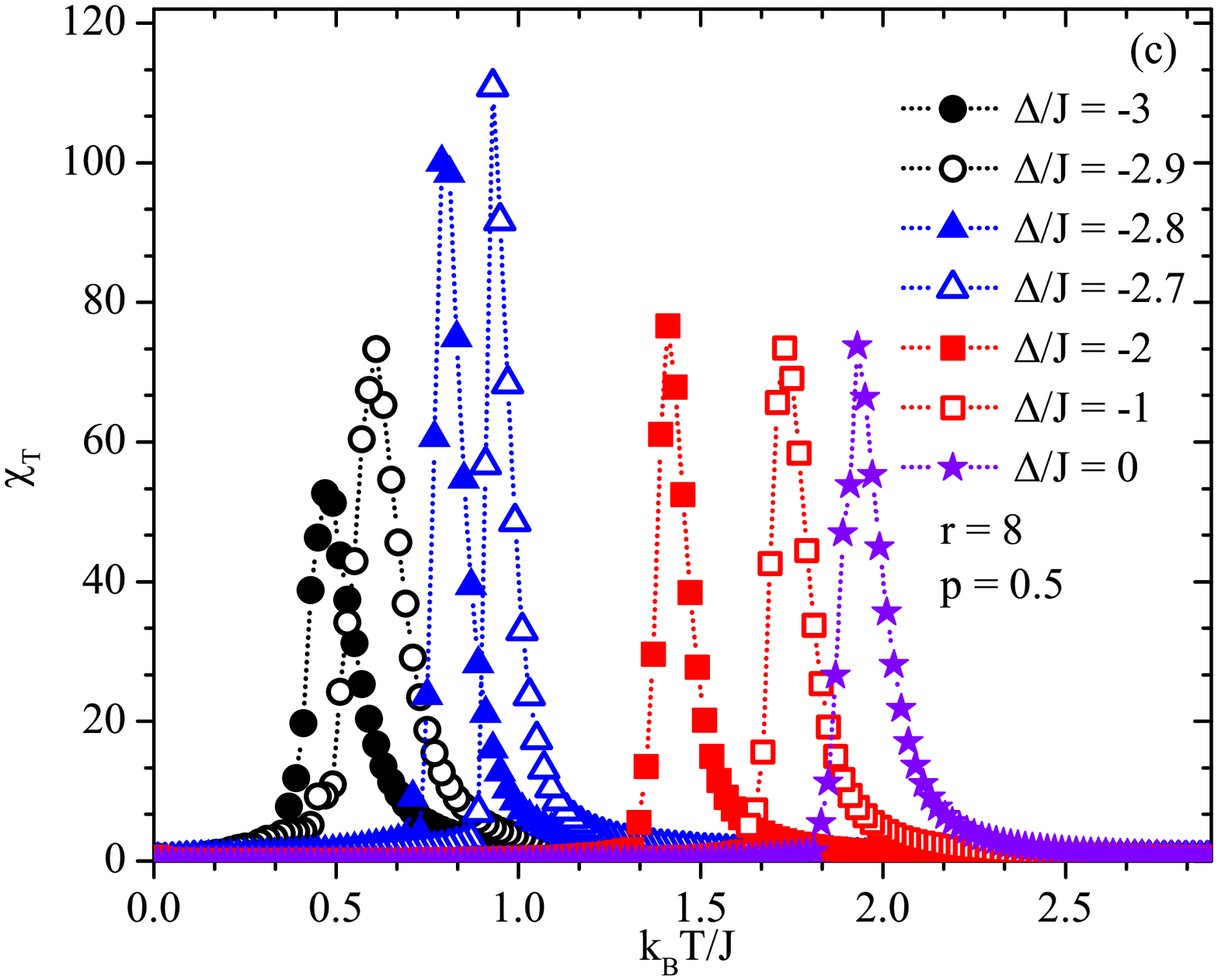}
\caption{(Color online) Magnetic phase diagrams in $\left(\Delta/J-k_{B}T_{C}/J\right)$
plane of the binary alloy spherical nanoparticle system, which are shown for different
values of the active concentrations of type-$A$ atoms such as $p=0.25, 0.5$ and $0.75$.
Influences of the single-ion anisotropies on the temperature dependencies of the total (b)
magnetizations $M_{T}$ and (c) susceptibility $\chi_{T}$ of the system with selected
value of $p=0.5$. All figures are presented for  value of radius $r=8$.
}\label{Fig4}
\end{figure*}

In the following analysis, we study the effects of the single-ion anisotropy term on the
ferromagnetic to  paramagnetic transition temperature of the binary alloy
nanoparticle system. In Figure \ref{Fig4}(a), the phase boundary curve in $\left(\Delta/J-k_{B}T_{C}/J\right)$
space in the case of three selected values of  concentration of type-$A$ atoms, $p=0.25, 0.5$ and $0.75$,
is presented for a nanoparticle radius of $r=8$. As one can see clearly from the figure that in the case
of low-negative and high-positive single-ion anisotropy region, the phase transition points saturate at particular
temperature values. This saturation temperature strongly depends on the concentration of spin-$1/2$ atoms. In the
high-positive $\Delta/J$ region, as one increases the number of $A$ atoms in the system, the saturation occurs
at low temperature values. This is due to the fact that replacing the spin-$1$ atoms with spin-$1/2$ atoms
in the system (increasing $p$) means decreasing the number of atoms under the influence of single-ion anisotropy
term and this causes the system to exhibit phase-transition at relatively low temperatures.
Conversely, for large negative $\Delta/J$ values, increasing $p$ results in saturation at high
temperature values. Apart from these, in small single-ion anisotropy region, critical
temperature rises when the amount of  $\Delta/J$ is increased.  Thermal variations of the total
magnetization and the corresponding magnetic susceptibility for the binary alloy spherical nanoparticle
system for several negative values of single-ion anisotropy terms and with a fixed concentration
value $p=0.5$ and nanoparticle radius $r=8$ are  shown in Figure \ref{Fig4}(b) and \ref{Fig4}(c),
respectively. For the present investigation, since the positive values of the $\Delta/J$ terms are no
significant effects on the magnetization profiles, except from its critical temperature,
we are restricted ourselves to the relatively low and high  negative $\Delta/J$ values.
As it is seen from Figure \ref{Fig4}(b) that the saturation value of the total
magnetization is $M_T=0.75$ when $\Delta/J=0$. However, because of the single-ion anisotropy term
acting on $B$-type atoms, as value of single-ion anisotropy decreases, the saturation value of the total
magnetization also decreases from $M_T=0.75$. Besides, the susceptibility peaks appear at lower
values of the temperature axis when the single-ion anisotropy term becomes
more negative. It should be underlined that we have not observed a
behavior indicating a first-order phase transition for all the selected
values of $p$  and single-ion anisotropy term.

\begin{figure*}[!h]
\center
\includegraphics[width=4.9cm]{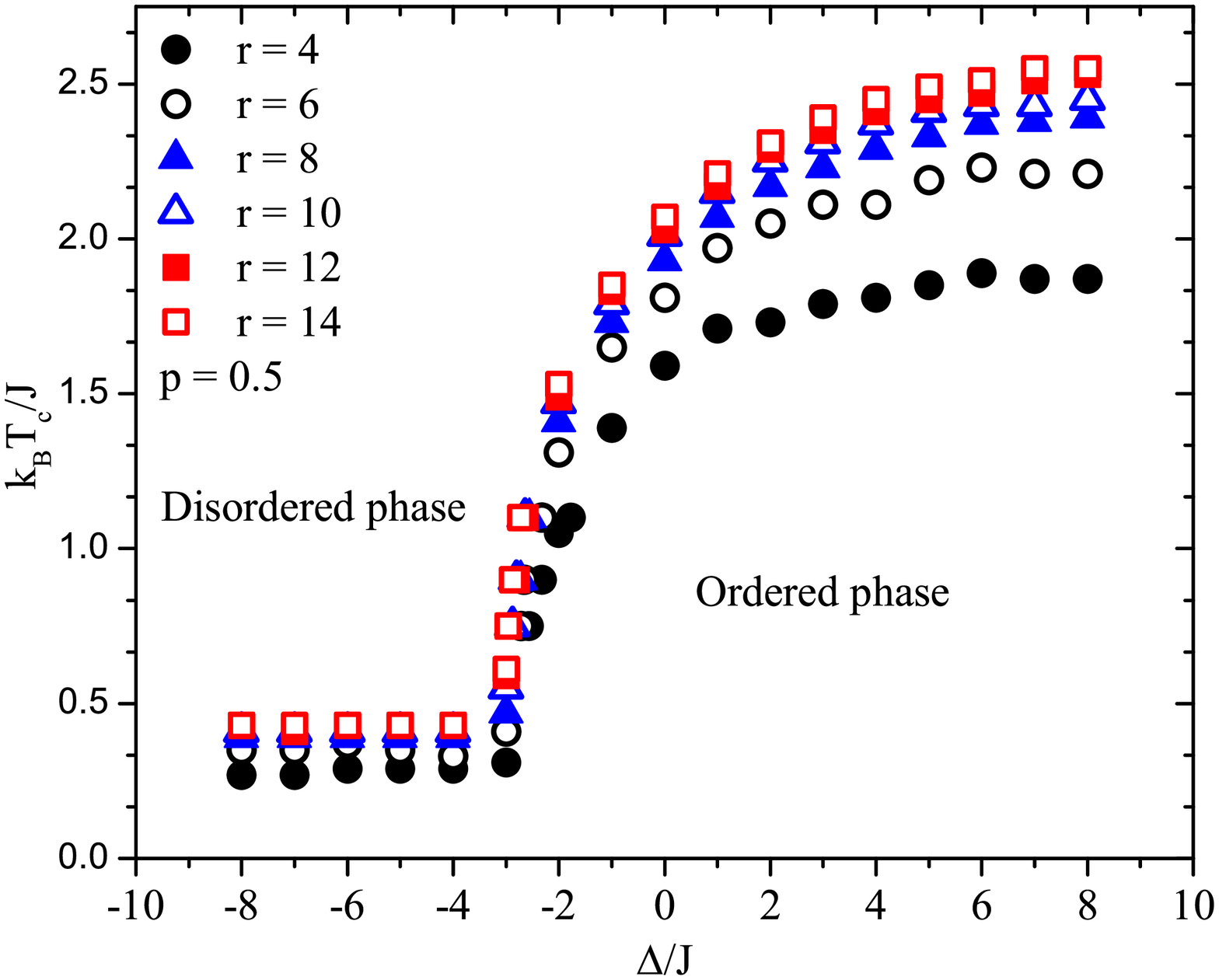}
\includegraphics[width=5.cm]{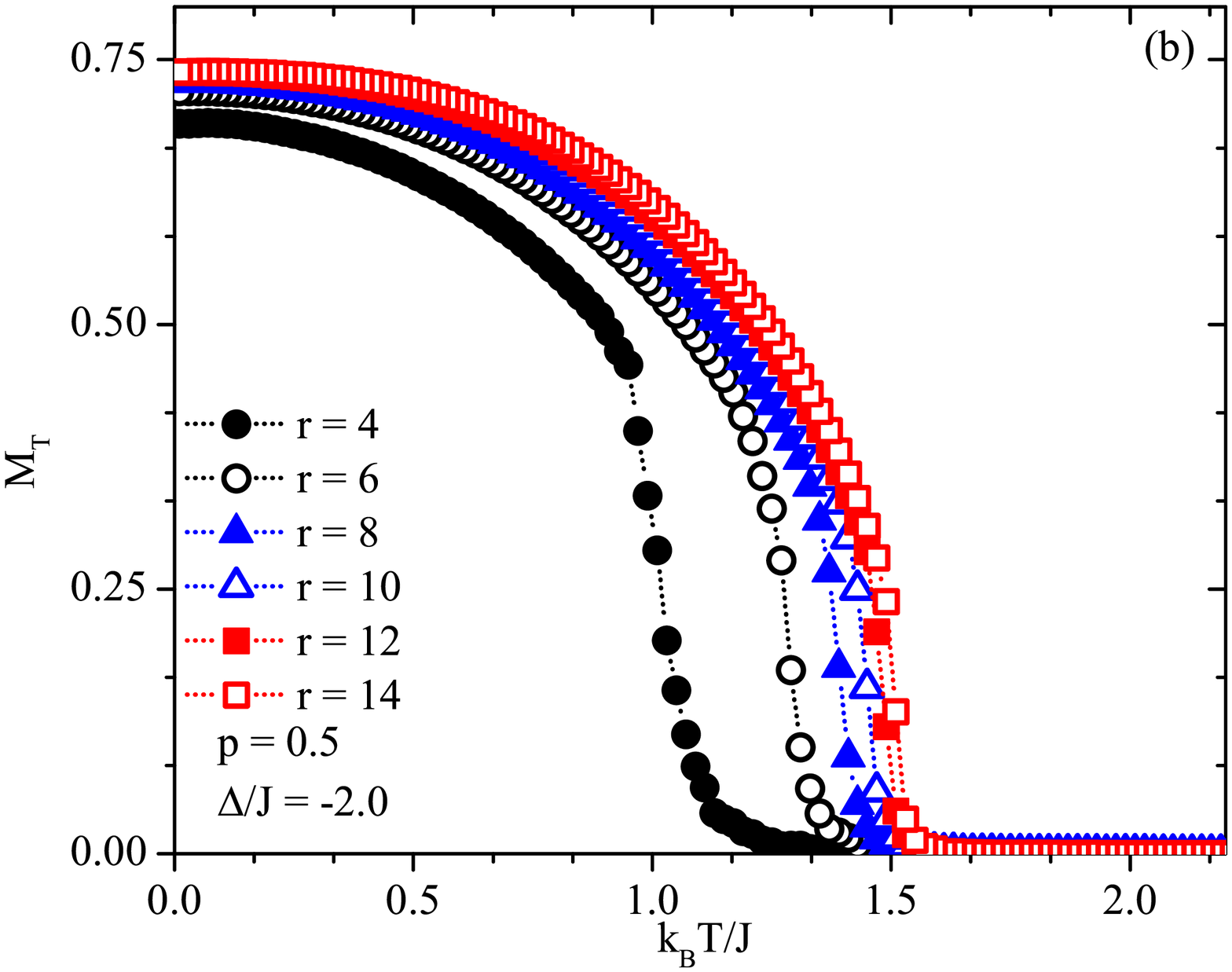}
\includegraphics[width=5.cm]{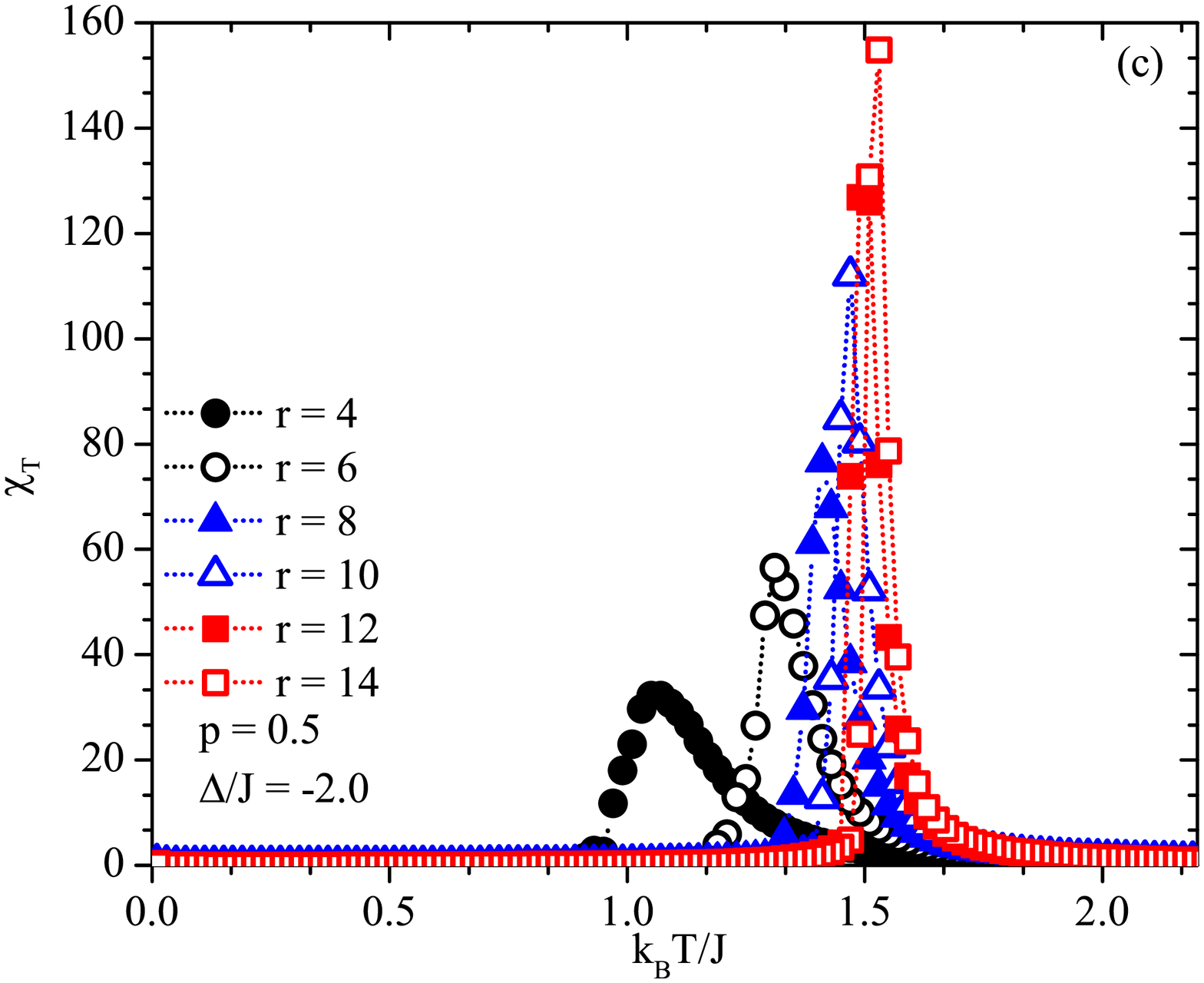}

\caption{(Color online) Effects of the varying values of radius of the binary alloy spherical nanoparticle on the magnetic
phase diagrams displayed in $(\Delta/J-k_{B}T_{C}/J)$ plane. Variations of the total (b) magnetizations $M_{T}$ and (c)
susceptibility $\chi_{T}$ as a function of the temperature for several values of radius of the system with $\Delta/J=-2$.
The curves are presented for a selected value of active concentration of  type-$A$ atoms,  $p=0.5$.} \label{Fig5}
\end{figure*}

Finally, we clarify the variations of the magnetic phase diagram in $\left(\Delta/J-k_{B}T_{C}/J\right)$
space with the size of the nanoparticle. Figure \ref{Fig5}(a) reveals the critical temperature versus
reduced single-ion anisotropy term for several values of  nanoparticle
radius, $r= 4, 6, 8, 10, 12$ and $14$.  It is possible to deduce from the figure that
the behavior of phase separation curve between ferromagnetic and paramagnetic phases is similar for
each value of $r$. One of the main differences is that  saturation value of the critical temperature
takes larger values with increasing particle size for all $\Delta/J$. In
figure \ref{Fig5}(b) and \ref{Fig5}(c), we give the temperature dependent of total magnetization and
susceptibility profiles of the system for several values of nanoparticle radius
with $p=0.5$ and $\Delta/J=-2$, respectively.  We note that the saturation
value of the total magnetization alters very little with varying
nanoparticle radius and becomes zero at different critical temperature values which
can be clearly seen from the positions of maximum values of the magnetic
susceptibility versus temperature curves.

\section{Conclusions}\label{Conclusion}
In this paper, we have considered the finite temperature phase transition features of a binary alloy spherical
nanoparticle with radius $r$ of the type $A_{p}B_{1-p}$. For this purpose, we use MC simulation method with
single-site update Metropolis algorithm. As a binary alloy spherical nanoparticle, we select a system consisting of
spin-$1/2$ and spin-$1$ magnetic components. They are randomly distributed depending on the selected value of the
active concentration of type-$A$ atoms. It is obvious that the cases of $p=1$ and $0$ correspond to the clean
Ising and Blume-Capel spherical nanoparticles, respectively. For both two values of $p$
mentioned above, there is no  lattice site containing disorder arising from the existence of
different species of magnetic components. However, the considered system composes of randomly located spin-$1/2$
and spin-$1$ components, except from these two values of $p$. Our findings clearly indicate that it is possible to
control the thermal and magnetic phase transition properties  of the system with the help of adjustable extrinsic
parameters $p$ and $r$ as well as other intrinsic system parameters. The most
important observations found in the present work can be shortly summarized as follows:

\begin{itemize}
  \item For a fixed value of $r$ and value of $\Delta/J=0$, our results suggest that
  magnetic phase transition point of the system prominently decreases when the value of active
  concentration of type-$A$ atoms increases starting from zero. One can see that
  phase transition curves given in $(p-k_{B}T_{C}/J)$ plane  exhibit an almost linear character,
  for all  considered values of $r$. It should be underlined that concentration dependencies of the
  critical temperatures of a binary alloy bulk system driven by a time dependent magnetic field
  have been discussed in Ref. \cite{Vatansever1} by means of MC simulation method.  If one
  compares the figure 2 of Ref. \cite{Vatansever1} with figure \ref{Fig2} of the present study, one can easily see the
  similarities and differences between the two systems.

  \item For a fixed value of $p$ and value of $\Delta/J=0$, it is found that when the radius of the
  spherical nanoparticle increases starting from $r=4$, the critical temperature begins to shift to higher
  temperature regions. If the $r$ further increases, and reaches a certain value which strongly depends on the Hamiltonian
  parameters, the phase transition point saturates a characteristic value.

  \item Moreover, thermal and magnetic phase transition features of the system sensitively depends on the
  single-ion anisotropy term. Both large negative and positive values of $\Delta/J$ term give rise to
  saturation of the critical temperature of the system to a characteristic value. It is worth to note that
  such type of observation originating from the existence of single-ion anisotropy
  has been found in  Ref. \cite{Vatansever2}. Here, dynamic phase transition properties of
 a binary alloy ferromagnetic alloy under the presence of a  time dependent magnetic field have been analyzed for a two
 dimensional square lattice.
\end{itemize}

Finally, we notice that it could be interesting to focus on the effects of the
different types of spin-spin interactions on the nonzero temperature phase transition properties of the
present system. Actually, such a binary alloy spherical nanoparticle can include four different values of spin-spin
interaction terms between considered spins, that is to say, $J_{AA}$, $J_{AB}$, $J_{BA}$ and $J_{BB}$ corresponding to the
randomly distributed components between $A-A$, $A-B$, $B-A$ and $B-B$, respectively.
It is also possible to improve the obtained results in this study by making use of a more realistic system such as
classic Heisenberg type Hamiltonian where the spins are allowed to point in any direction.
From theoretical perspective, this type of model containing the  different types of spin-spin interactions
mentioned above may be subject of a future work in order to provide deeper understanding of thermal
and magnetic phase transition properties of the  nanoparticle systems.

\section*{Acknowledgements}
The numerical calculations reported in this paper were
performed at T\"{U}B\.{I}TAK ULAKB\.{I}M (Turkish agency), High Performance and
Grid Computing Center (TRUBA	 Resources).


\begin{thebibliography}{99}

\bibitem{Berkowitz} A.E. Berkowitz, R.H. Kodama, S.A. Makhlouf, F.T. Parker, F.E. Spada, E.J. McNiff Jr.,
S. Foner, J. Magn. Magn. Mat. 196 (1999) 591.
\bibitem{Plumer} M.L. Plumer,   J. van Ek, D. Weller, The Physics of Ultrahigh-
Density Magnetic Recording(Springer Series in Surface Sciences) 41 (2001).
\bibitem{Hayashi} T. Hayashi, S. Hirono,  M. Tomita,  S. Umemura, Nature 381 (1996) 772.
\bibitem{Comstock} R.L. Comstock, Introduction to Magnetism and Magnetic Recording, Wiley, New York (1999).
\bibitem{Pankhurst} Q.A. Pankhurst, J. Connolly, S.K. Jones, J. Dobson, J. Phys. D: Appl.
Phys. 36  (2003) R167.
\bibitem{Rivas} J. Rivas,  M. Bo\~{n}obre-L\'{o}pez,  Y. Pi\~{n}eiro-Redondo,
B. Rivas,   M.A. L\~{o}pez-Quintela, J. Magn. Magn. Mater. 324  (2012) 3499.
\bibitem{Blakemore} R.P. Blakemore, Annu. Rev. Microbiol. 36   (1982) 217.
\bibitem{Bazylinski} D.A. Bazylinski, R.B. Frankel, H.W.  Jannasch,  Nature 334 (1988) 518.
\bibitem{Ivanov} Y.P. Ivanov, A. Alfadhel, M. Alnassar, J.E. Perez, M. Vazquez, A. Chuvilin, J. Kosel,
Sci. Rep. 6 (2016) 24189.
\bibitem{Bhatt} P. Bhatt, A. Kumar, S.S. Meena, M.D. Mukadam, S.M. Yusuf, Chem.  Phys. Lett. 651 (2016) 155.
\bibitem{Leite} V.S. Leite, W. Figueiredo, Physica A 350 (2005) 379.
\bibitem{Kaneyoshi1} T. Kaneyoshi, J. Magn. Magn. Mater. 321 (2009) 3430.
\bibitem{Kaneyoshi2} T. Kaneyoshi, Physica Status Solidi B 246 (2009) 2359.
\bibitem{Garanin} D.A. Garanin, H. Kachkachi, Phys. Rev. Lett. 90 (2003) 065504.
\bibitem{Wang} H. Wang, Y. Zhou, D.L. Lin, C. Wang, Physica Status Solidi B 232 (2002) 254.
\bibitem{Kocakaplan} Y. Kocakaplan, M. Keskin, J. Appl. Phys. 116 (2014) 093904.
\bibitem{Jiang1} W. Jiang, X.-X. Li, L.-M. Liu, J.-N. Chen, F. Zhang, J. Magn. Magn. Mater. 353 (2014) 90.
\bibitem{Kaneyoshi3} T. Kaneyoshi, J. Phys. Chem. Solids 96-97 (2016) 1.
\bibitem{Kaneyoshi4} T. Kaneyoshi, Solid State Commun. 244 (2016) 51.
\bibitem{Bouhou} S. Bouhou, I. Essaoudi, A. Ainane, R. Ahuja, Physica B 481 (2016) 124.
\bibitem{Kantar} E. Kantar, J. Alloys Compd.  676 (2016) 337.
\bibitem{Hamri} M. El Hamri, S. Bouhou, I. Essaoudi, A. Ainane, R. Ahuja,
 F. Dujardin, Appl. Phys. A 122 (2016) 202.
\bibitem{Magoussi} H. Magoussi, A. Zaim, M. Kerouad, Solid State Commun. 200 (2014) 32.
\bibitem{Margaris} G. Margaris, K.N. Trohidou, J.J. Nogu\'{e}s, Adv. Mater. 24 (2012) 4331.
\bibitem{Vasilakaki} M. Vasilakaki, K. Trohidou, J. Nogu\'{e}s, Sci. Rep. 5 (2015) 9609.
\bibitem{Drissi} L.B. Drissi, S. Zriouel, J. Stat. Mech. (2016)  053206.
\bibitem{Zaim1} A. Zaim, M. Kerouad, Physica A 389 (2010) 3435.
\bibitem{Zaim2} N. Zaim, A. Zaim, M. Kerouad, J.  Alloys Compd. 663 (2016) 516.
\bibitem{Zaim3} N. Zaim, A. Zaim, M. Kerouad, Solid State Commun. 246 (2016) 23.
\bibitem{Aouini} S. Aouini, S. Ziti, H. Labrim, L. Bahmad, Solid State Commun.  241 (2016) 14.
\bibitem{Feraoun} A. Feraoun, A. Zaim, M. Kerouad, J. Phys. Chem. Solids 96-97 (2016) 75.
\bibitem{Jiang2} W. Jiang, J.-Q. Huang, Physica E 78 (2016) 115.
\bibitem{Russier} V. Russier, J. Magn. Magn. Mater. 409 (2016) 50.

\bibitem{Thorpe} M.F. Thorpe, A.R. McGurn, Phys. Rev. B 20 (1979) 2142.
\bibitem{Tahir} R.A. Tahir-Kheli, T. Kawasaki, J. Phys. C 10 (1977) 2207.
\bibitem{Plascak} J.A. Plascak, Physica A 198 (1993) 655.
\bibitem{Katsura} S. Katsura, F. Matsubara, Can. J. Phys. 52 (1974) 120.
\bibitem{Honmura} R. Honmura, A.F. Khater, I.P. Fittipaldi, T. Kaneyoshi, Solid State
Commun. 41 (1982) 385.
\bibitem{Kaneyoshi5} T. Kaneyoshi, Phys. Rev. B 34 (1986) 7866.
\bibitem{Kaneyoshi6} T. Kaneyoshi, Phys. Rev. B 33 (1986) 7688.
\bibitem{Kaneyoshi7} T. Kaneyoshi, Z.Y. Li, Phys. Rev. B 35 (1869).
\bibitem{Kaneyoshi8} T. Kaneyoshi, Phys. Rev. B 39 (1989) 12134.
\bibitem{Scholten1} P.D. Scholten, Phys. Rev. B 32 (1985) 345.
\bibitem{Scholten2} P.D. Scholten, Phys. Rev. B 40 (1989) 4981.
\bibitem{Godoy} M. Godoy, W. Figueiredo, Int. J. Mod. Phys. C 20 (2009) 47.
\bibitem{Cambui} D.S. Cambui, A.S. De Arruda, M. Godoy, Int. J. Mod. Phys. C 23 (2012) 1240015.

\bibitem{Binder}  K. Binder, Monte Carlo Methods in Statistical Physics (Springer, Berlin, 1979).
\bibitem{Newman} M.E.J. Newman, G.T. Barkema, Monte Carlo Methods in Statistical Physics (Oxford University Press, New York, 1999).
\bibitem{Yuksel} Y. Y\"{u}ksel, Phys. Rev. E 91 (2015) 032149.
\bibitem{YunBin} L. YunBin, L. ShuZhi, X. Bin, C. Jia, P. HaoJun, Z. Chun, Z. HuiYing,
X. HaoWen, O. YiFang, Z. BangWei, Science China Physics, Mechanics and Astronomy, 54 (2011) 897,
and references therein.
\bibitem{Vatansever1} E. Vatansever, H. Polat, Phys. Lett. A 379 (2015) 1568.
\bibitem{Vatansever2} E. Vatansever, U. Akinci, H. Polat, J. Magn. Magn. Mater., 389 (2015) 40.






\end{thebibliography}
\end{document}